\documentclass[manuscript]{aastex}
\usepackage{emulateapj5}
\usepackage{apjfonts}
\usepackage{epsfig}

\def\oiii{{[O \sc iii]}\,}

\def\nii{{[N \sc ii]}\,}
\def\sii{{[S \sc ii]}\,}

\slugcomment{Received 2009 July 27; accepted 2009 September 30}
\journalinfo{The Astrophysical Journal Letters in Press}

\begin{document}

\title{Active galactic nuclei with double-peaked narrow lines: are they dual AGNs?}
\author{Jian-Min Wang\altaffilmark{1,2}, 
Yan-Mei Chen\altaffilmark{1,3}, 
Chen Hu\altaffilmark{1}, 
Wei-Ming Mao\altaffilmark{4},\\ 
Shu Zhang\altaffilmark{1} and
Wei-Hao Bian\altaffilmark{5}}
\vglue 0.1cm

\altaffiltext{1}{
Key Laboratory for Particle Astrophysics, Institute of High Energy Physics,
Chinese Academy of Sciences, 19B Yuquan Road, Beijing 100049, China
}
\altaffiltext{2}{
Theoretical Physics Center for Science Facilities, Chinese Academy of Sciences, China
}
\altaffiltext{3}{
Graduate University of Chinese Academy of Sciences, 19A Yuquan Road, Beijing 100049, China
}

\altaffiltext{4}{
National Astronomical Observatories, Chinese Academy of Sciences, 20 Datun Road, Beijing 100020,
China
}

\altaffiltext{5}{Physics Department and Institute of Theoretical Physics, Nanjing Normal University, 
Nanjing 210097, China
}

\begin{abstract}
Double-peaked \oiii\, profiles in active galactic nuclei (AGNs) may provide evidence
for the existence of dual AGNs, but a good diagnostic for selecting them is currently lacking. 
Starting from $\sim$ 7000 active galaxies in SDSS DR7, we assemble a sample of 87 type 2 AGNs 
with double-peaked \oiii\, profiles. The nuclear obscuration in the type 2 AGNs allows us to 
determine redshifts of host galaxies through stellar absorption lines. We typically find that 
one peak is redshifted and another is blueshifted relative to the host galaxy. We find a strong 
correlation between the ratios of the shifts and the double peak fluxes. The correlation can be 
naturally explained by the Keplerian relation predicted by models of co-rotating dual AGNs. The 
current sample statistically favors that most of the \oiii\ double-peaked sources are dual AGNs 
and disfavors other explanations, such as rotating disk and outflows. These dual AGNs have a 
separation distance at $\sim 1$ kpc scale, showing an intermediate phase of merging systems.
The appearance of dual AGNs is about $\sim 10^{-2}$, impacting on the current observational 
deficit of binary supermassive black holes with a probability of $\sim 10^{-4}$ (Boroson 
\& Lauer). 
\end{abstract}

\section{Introduction}
Supermassive black hole binaries and dual AGNs are the natural result of the current $\Lambda$CDM 
cosmological paradigm, in which galaxies grow hierarchically through minor or major mergers. However, 
evidence for such systems is elusive and confirmed cases are rare despite circumstantial evidence. 
Examples can be found in Kochanek et al. (1999), Junkkarinen et al. (2001), Ballo et al. (2004), 
Maness et al. (2004), Guainazzi et al. (2005), Rodriguez et al. (2006), Hudson et al. (2006), Barth 
et al. (2008). The best-known examples of dual AGNs are found in NGC 6240 (Komossa 
et al. 2003), EGSD2 J142033+525917 (Gerke et al. 2007), EGSD2 J141550+520929 (Comerford et al. 2009a), 
the $z=0.36$ galaxy COSMOS J100043+020637 (Comerford et al. 2009b). 
Ideally, analysis of a large statistical sample should be undertaken.

The narrow emission lines in AGNs are generally thought to be produced by clouds in the narrow line
regions at a scale of $\sim 1$ kpc, where they are photoionized by the central energy source (e.g.
Netzer et al. 2006). It is common that the \oiii$\lambda$5007 line has an asymmetric blue wing 
(Boroson 2005, Komossa et al. 2008), which could be caused by winds and outflows. On the other hand, 
about 1\% AGNs have double-peaked \oiii\, profiles. The origin of the double-peaked emission lines 
remains unclear, but they have been interpreted as evidence for bi-polar outflows in the early 1980s 
(Heckman et al. 1981, 1984; Whittle 1985a,b,c)
or disk-shaped narrow line regions (Greene et al. 2005). Evidence based on objects mentioned
above suggests dual AGNs as an alternative explanation of the double-peaked
profiles. However, more robust observational evidence is needed.

In this Letter, we select type 2 AGNs from SDSS DR7. The obscuration of the nucleus in these sources 
allows us to determine the redshifts of the host galaxies through the stellar absorption lines and
investigate the properties of the double peaks. We find a strong correlation between the ratios 
of the shifts and the double peak fluxes. This is most naturally explained by the Keplerian 
relation of co-rotating dual AGNs.

\section{Sample and data analysis}
\subsection{Sample selection}
We start from the galaxy sample of SDSS DR7 and select active galaxies based 
on the widely used BPT diagram (Baldwin et al. 1981). 
Using the measured fluxes of narrow emission lines from the MPA/JHU
catalog\footnote{The MPA/JHU catalog can be downloaded from http://www.mpa-garching.mpg.de/SDSS/DR7.},
we separate type 2 AGNs from other sources according to their line ratios:
$\log ($\oiii/H$\beta) >0.61/\{\log ($\nii/H$\alpha)-0.47\} +1.19$ (the solid curve in Fig. 2 from 
Kewley et al. 2001), or $\log ($\nii/H$\alpha)>0.47$.
We select those objects with SDSS $r-$band magnitude $r<17.7$, redshift $z \le 0.15$ and
equivalent width of \oiii$\lambda5007$ greater than 5. In total, 6780 type 2 AGNs follow
from this criterion. We use stellar population synthesis models to separate the spectra of
the host galaxies (Tremonti et al. 2004) from those of the AGNs. The redshifts of the host 
galaxies are re-measured through the absorption lines (see \S2.2 for details).
\oiii$\lambda$5007 is fitted by two Gaussian profiles in order to select the double-peaked
AGNs. Those sources having two \oiii$\lambda5007$ components (one blueshifted and one redshifted, 
relative to the host galaxies) separated by 
$\Delta\lambda \ge 1 \rm{\AA}$ (based on the redshift accuracy, see \S2.2) and a 
flux ratio in the range of $\sim [0.1,10]$, are selected. This results in 286 candidates. 
We visually inspect their spectra and remove those galaxies with a false component. For these 
galaxies which have very large or small flux ratios, the double-peaked qprofiles
are not convincing. Finally, we are left with 87 sources having a flux ratio in the range of 
[0.3, 3], 44 of which have high enough S/N of the spectra to see similar double-peaked structures 
in other lines (e.g. \oiii$\lambda4959$, H$\beta$, H$\alpha$, \nii, \sii). Fig. 1 shows one example.

\begin{center}
\includegraphics[width=3.1in]{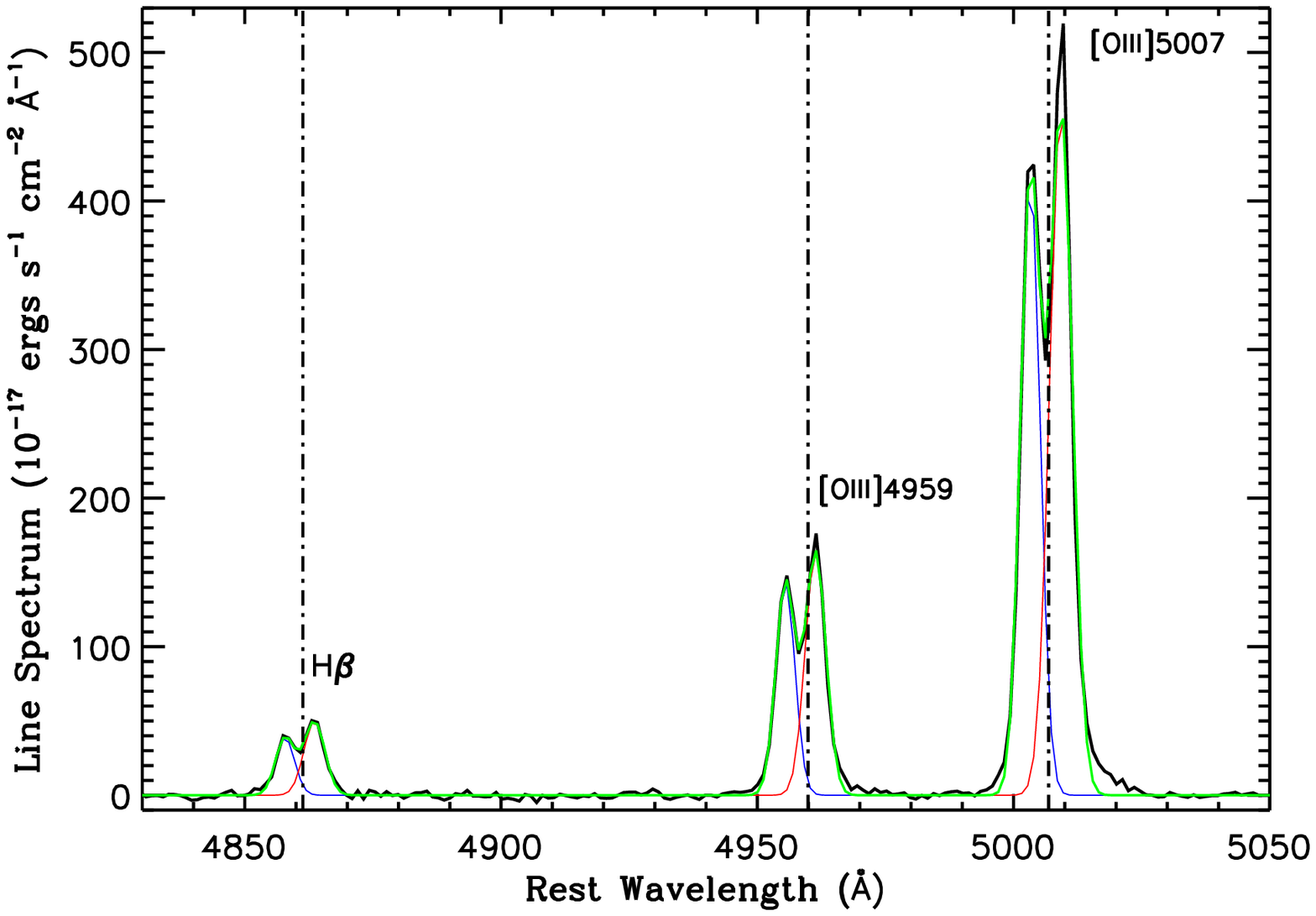}
\includegraphics[width=3.1in]{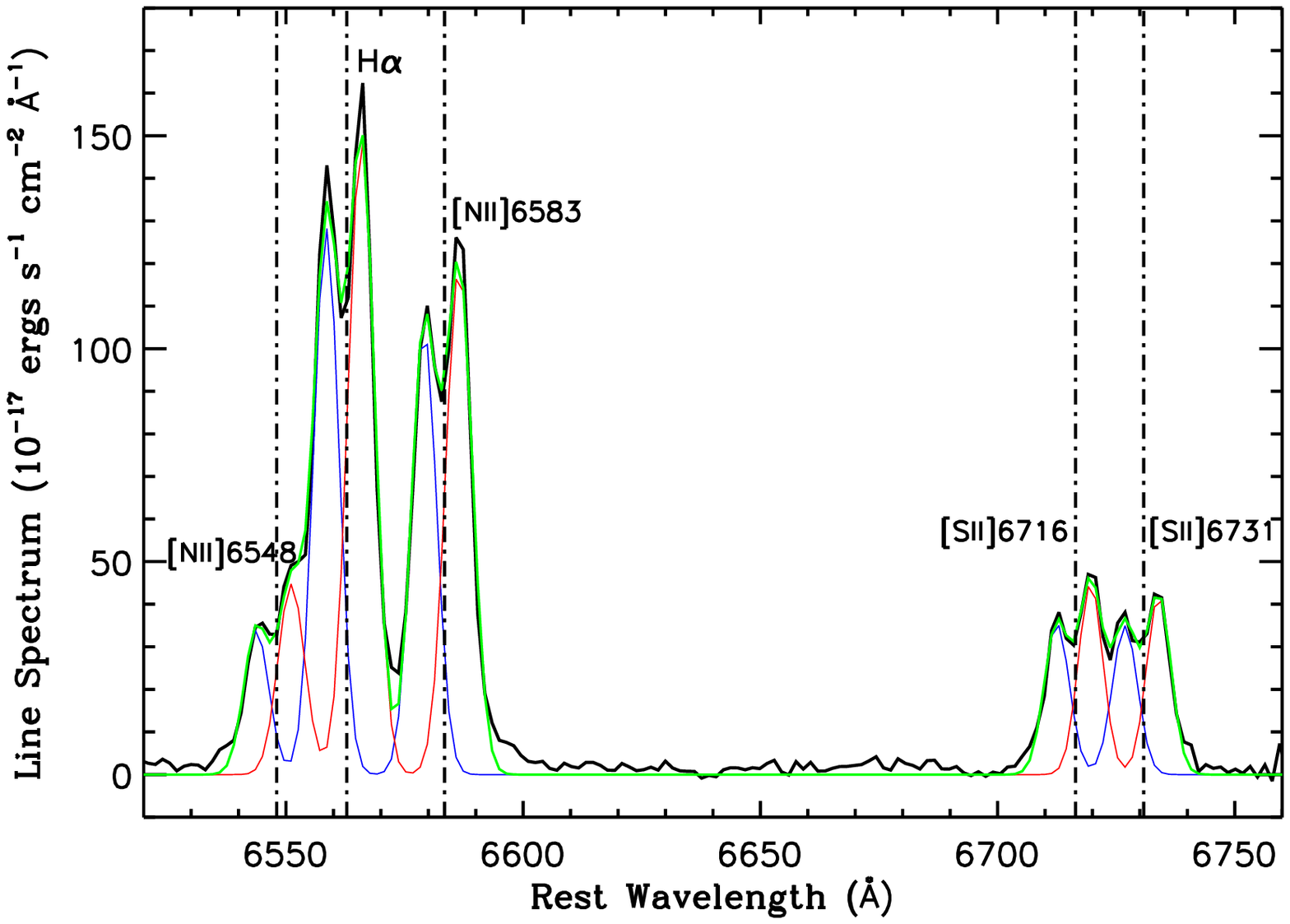}
\figcaption{Spectra of object SDSS J080218.65+304622.8 selected from our sample. Blue lines are the blue-shifted 
components, and red lines are the redshifted components. The green line represents the final fit.
It is generally found that the separation ratios of the two components of H$\beta$, H$\alpha$, \nii, \sii
are roughly equal to that of [O {\sc iii}]. This indicates that the source is a candidate
for a dual AGN.}
\end{center}
\vglue 0.1cm

\subsection{Determinations of galaxy redshifts}
One key parameter in this work is the ratio of wavelength shifts of the two components relative to 
the host galaxy. This ratio is sensitive to the host galaxy redshift. In order to get a robust shift 
ratio, a host galaxy redshift accuracy of $\Delta z/z=\Delta \lambda/\lambda\sim 2 \times 10^{-4}$ 
is required based on the redshift range of our sample and 1\AA\ separation limit in term of SDSS
spectral resolution. We first fit the 
galaxy spectrum by a stellar population model (Tremonti et al. 2004), 
under the basic assumption that any galaxy star formation history can be approximated by a sum of
discrete bursts. The library of template spectra is composed of single stellar population 
models generated using the population synthesis code of BC03 (Bruzual \& Charlot 2003), including 
models of ten different ages (0.005, 0.025, 0.1, 0.2, 0.6, 0.9, 1.4, 2.5, 5, 10 Gyrs) and four 
metallicities (0.004, 0.008, 0.017, 0.05). The template spectra are convolved to the 
stellar velocity dispersion measured for each SDSS galaxy, and four best fitting model 
spectra with fixed metallicity are constructed from a non-negative linear combination 
of the template spectra, with dust attenuation modeled as an additional free parameter. 
The results of the stellar population synthesis can be found from the SDSS-MPA webpage. We
use the redshift given by SDSS pipeline as initial value for our iterative fitting technique.
The absorption line redshift is determined by the fit having the lowest $\chi^2$ value.
The uncertainty of the redshift is given by the width of the $\chi^2$ 
minimum at which $\Delta \chi^2=1$. We give the redshifts of the host galaxies, the blue and 
redshifts of the two components of narrow emission lines and their corresponding fluxes
in Table 1, respectively. 

\section{Properties of the double-peaked lines}
\subsection{The two components in the BPT diagram}
The location of AGNs in the BPT diagram is mainly determined by the SED of the photoionizing
central engine. Since SEDs are strongly controlled by the Eddington ratios
(Wang et al. 2004; Shemmer et al. 2006; Yang, et al. 2007; Gu \& Cao 2009), 
the separation of the two components (${\cal D}$) in the BPT diagram provides quantitative
information on the degree of similarity between the Eddington ratios of accreting black holes. 

{\centering
\figurenum{2}
\includegraphics[scale=0.55,angle=0]{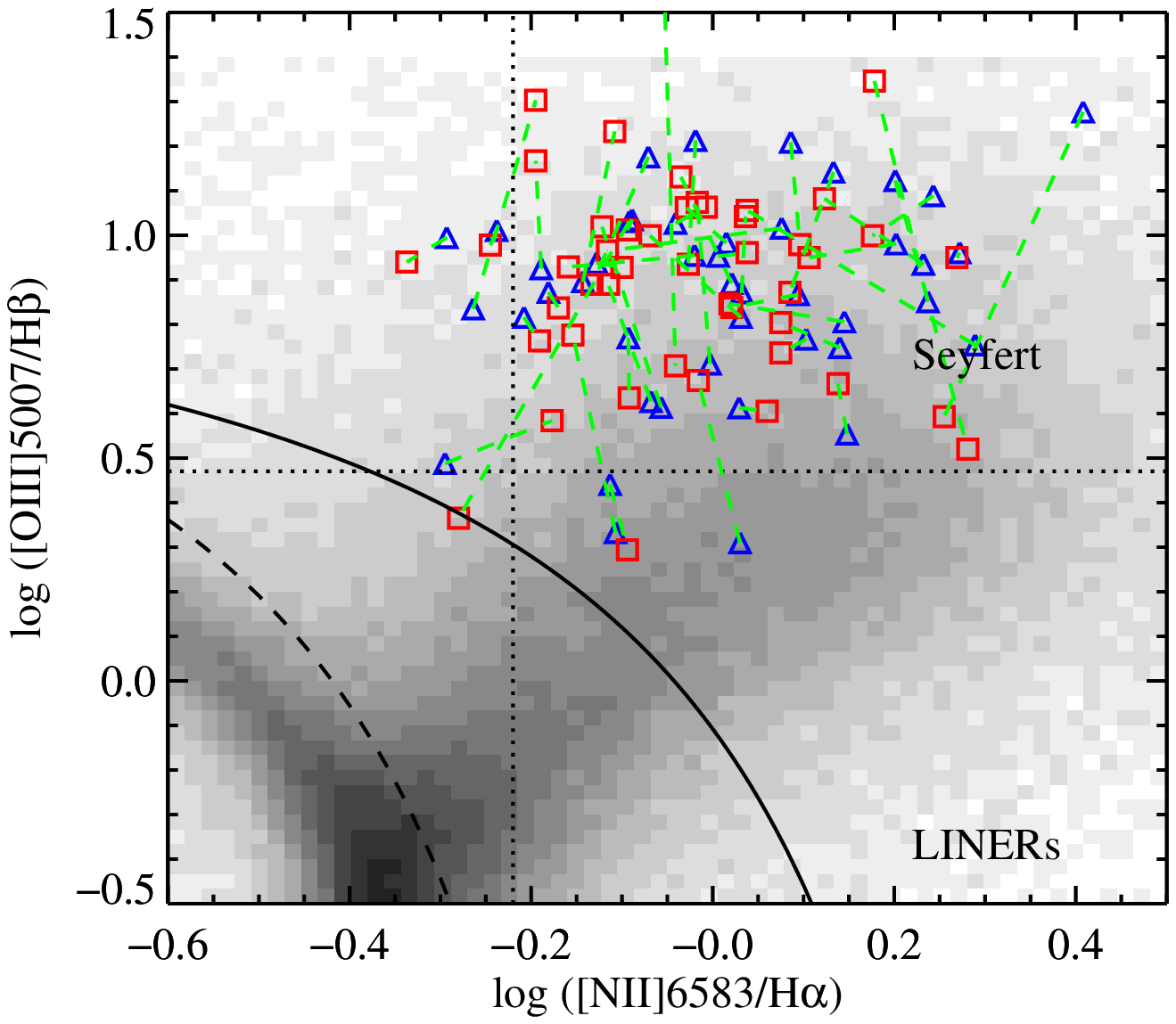}
\figcaption{The BPT diagram of each of the two components. The solid and dashed lines correspond 
to the dividing lines between AGN and star forming galaxies of Kewley et al. (2001) and 
Kauffmann et al. (2003), respectively. Seyfert galaxies are often defined to have 
\oiii/H$\beta>3$ (horizontal dotted line) and \nii/H$\alpha>0.6$ (vertical dotted line), 
and LINERs to have \oiii/H$\beta<3$ and \nii/H$\alpha<0.6$. Blue triangles indicate the 
buleshifted components whereas the red squares indicate the redshifted ones. The two 
components belonging to a single source are linked by dashed green lines. The double-peaked components 
in each object are very similar in most of the sample, indicating that 
the two components are produced by similar illuminating sources. Overplotted is the general population 
of SDSS DR7 galaxies.}
\label{fig2}
}
\vglue 0.3cm

We examine the locations of the two components in the BPT diagram in order to explore their
central engines.
There are 44 objects in our sample with clear double-peaked structure in H$\beta$, \oiii,
H$\alpha$ and \nii. Fig. 2 shows the two components in the BPT diagram for the 44 AGNs. 
The solid and dashed lines correspond to the dividing lines between AGN and star forming 
galaxies of Kewley et al. (2001) and Kauffmann et al. (2003), respectively. The entire 
sample of SDSS DR7 galaxies is overplotted for comparison. Blue triangles indicate the 
blue-shifted components, while red squares indicate the red-shifted components. Green 
dashed lines link the two components in 
each object. From Fig. 2, we can learn that each component of all the 44 objects are 
located in the Seyfert regime. The ${\cal D}-$distribution is peaked around zero from 
Fig. 3, rather than flat. The mean value of ${\cal D}$ is about 0.23.
This shows that for each source, the two components have similar 
Eddington ratios.

We would like to point out that it is beyond the scope of the present paper to investigate 
the qualitative relation between ${\cal D}$ and the degree of similarity between 
the two emitters. However it is robust that the smaller the ${\cal D}$, the more similar the 
two emitters are. We have to note that no object has ${\cal D}=0$, which could mean that
the two narrow line regions share same or one illuminating source.

{\centering
\figurenum{3}
\includegraphics[scale=0.45,angle=0]{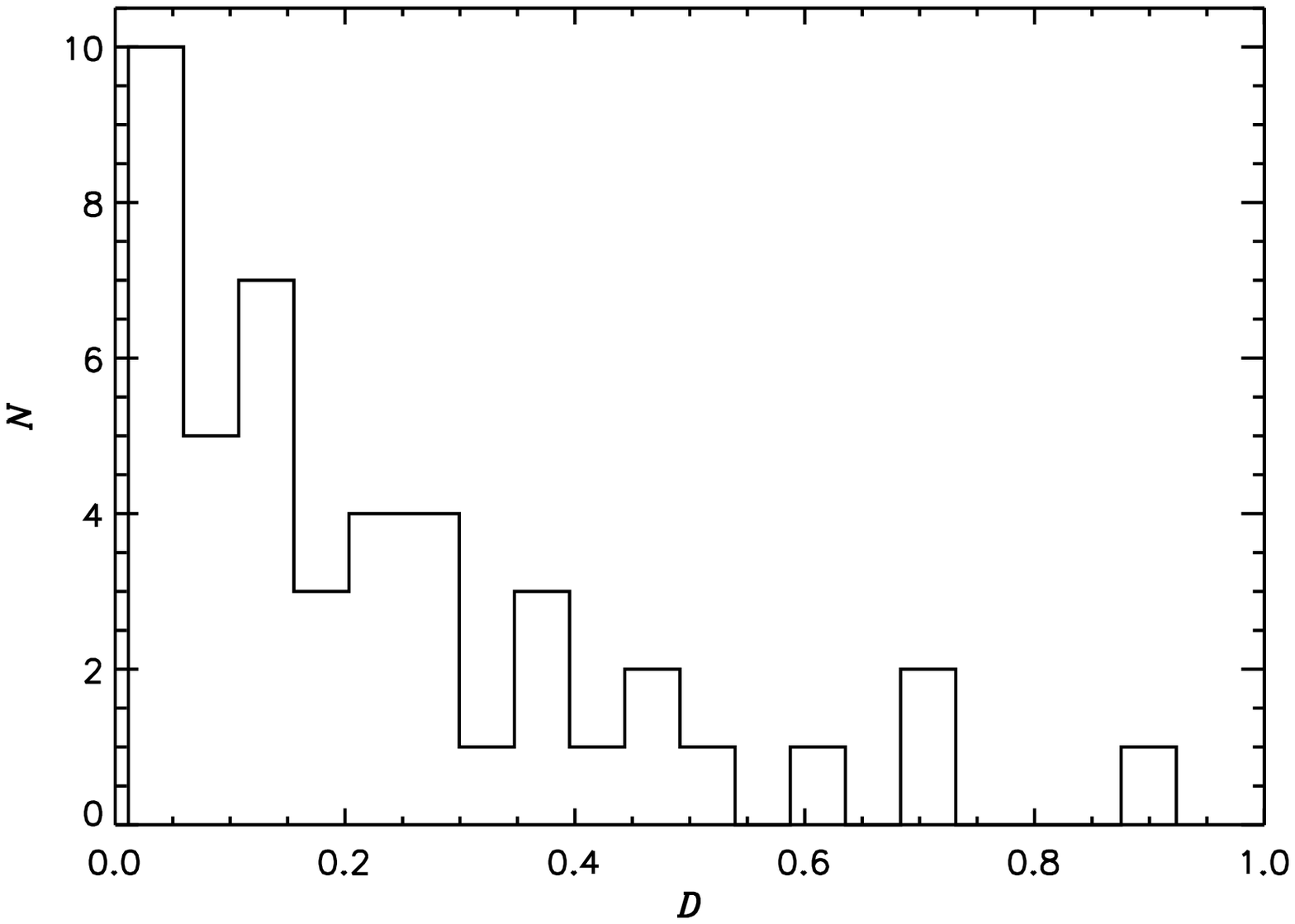}
\figcaption{The ${\cal D}-$distribution. The sources peak around zero. The mean value 
$\langle{\cal D}\rangle\approx 0.23$. This implies that the two components have similar 
photoionizing energy sources.}
\label{fig3}
}
\vglue 0.2cm

\subsection{A correlation between the line shifts and the luminosities}
In order to further study the properties of the double-peaked \oiii\  and its origin,
we measure the shifts $\Delta \lambda_1=\left|\lambda_1-\lambda_0\right|$ and 
$\Delta \lambda_2=\left|\lambda_2-\lambda_0\right|$, where $\lambda_1$ and $\lambda_2$ 
are the peak wavelength of the blue and red components of \oiii$\lambda5007$ in the rest 
frame of the host galaxies, respectively, and $\lambda_0=5006.84$\AA, and their corresponding 
$L_{[\rm O~III],1}$ and $L_{[\rm O~III],2}$ are the corresponding luminosities of the 
blue and red components. We 
define $\Delta \lambda_1/\Delta \lambda_2$ and $L_{[\rm O~III],1}/L_{[\rm O~III],2}$ as 
the ratios of the shifts and double peak fluxes, respectively. Fig. 4 shows a striking 
correlation between 
$\Delta \lambda_1/\Delta \lambda_2$ and $L_{[\rm O~III],1}/L_{[\rm O~III],2}$
which we fit as 
\begin{equation}
\log\left(\frac{L_{[\rm O~III],1}}{L_{[\rm O~III],2}}\right)
    =(0.005\pm 0.010)-(0.960\pm 0.11)\log\left(\frac{\Delta \lambda_1}{\Delta \lambda_2}\right),
\end{equation}
with the Pearson correlation coefficient of $-$0.43 and a chance probability of $3\times 10^{-5}$. 
We would like to stress here that this correlation poses a strong constraint on the
origin of the double peaks since it establishes a physical relation between the two
components. In the following section, we discuss the possible 
origin of the double-peaked line structure according to this correlation.

\section{Discussion}
The origin of double-peaked Balmer broad lines in quasars is likely ascribed to the outer
part of the accretion disk surrounding an SMBH (Eracleous et al. 2003), but only one potential 
exception from 17500 SDSS DR5 quasars (Boroson \& Lauer 2009). The size of narrow line regions 
scales with \oiii$\lambda5007$ luminosity as $R_{\rm NLR}=1.2 L_{\rm [O~III],42}^{0.5}$kpc, 
where $L_{\rm [O~III],42}=L_{\rm [O~III]}/10^{42}$ erg s$^{-1}$ (Netzer 
et al. 2006). Double-peaked narrow emission lines of active galaxies are plausibly caused by 
two orbiting narrow line regions of dual AGNs during an ongoing merger of their host galaxies 
(Bogdanovic et al. 2008), but it is difficult to distinguish such a configuration from the 
occasional projections of quasars, gravitational lensing effects (Mortlock et al. 1999; 
Kochanek et al. 1999), disks, bi-polar outflows and jet/cloud interactions (Heckman et al. 
2009). We first consider the scenario of dual AGNs as explanation of the strong correlation 
found in \S3.2. 

{\centering
\figurenum{4}
\includegraphics[scale=0.6,angle=0]{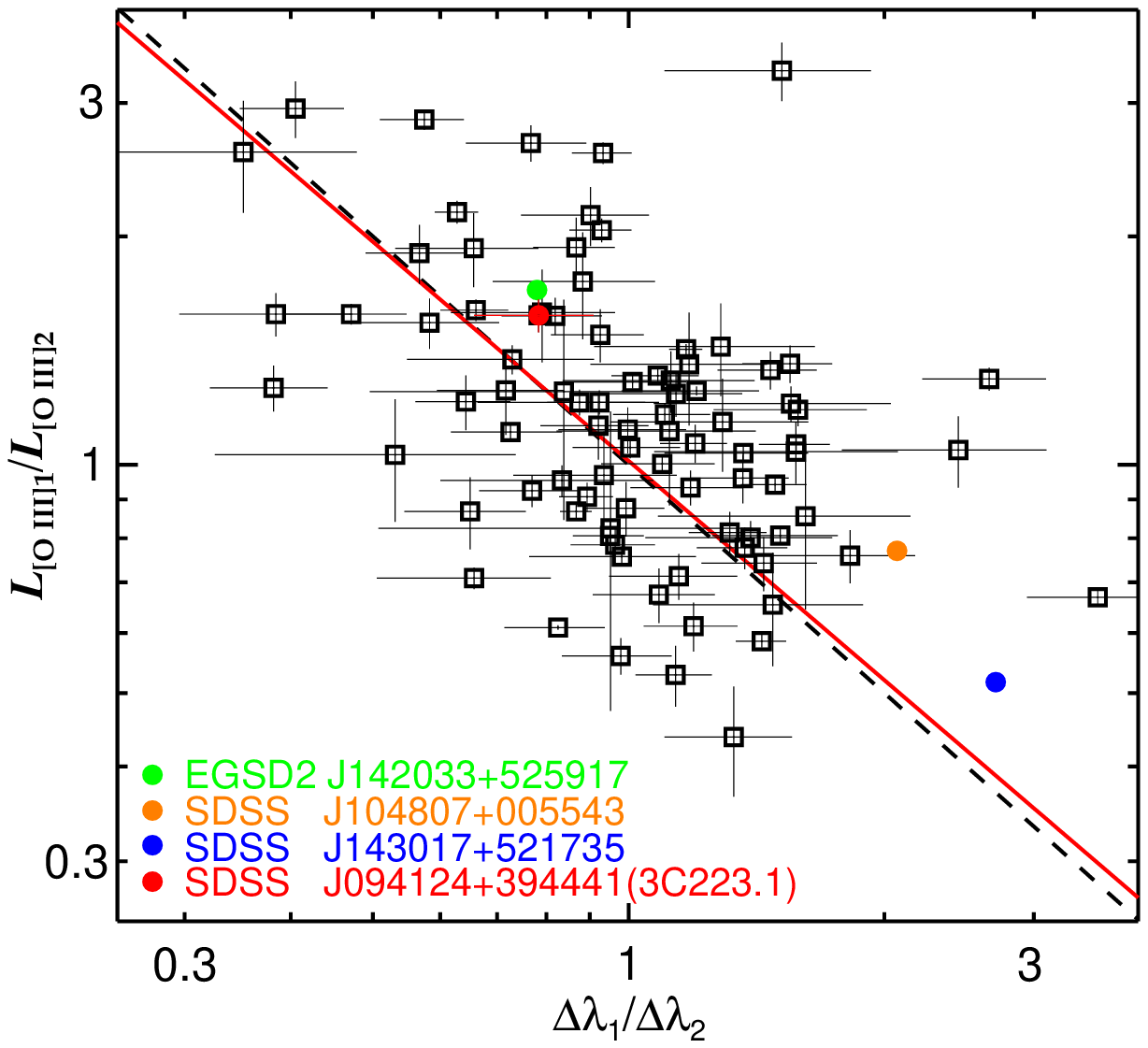}
\figcaption{
Correlation between the flux and shift ratios of the two \oiii$\lambda5007$ components. Black 
squares represent our sample. The red solid line is the best-fit linear correlation given in 
eq. (1). The dashed line is eq. (5) by setting $\epsilon_{1,2}=1$. We find that the Keplerian
relation matches the observed correlation very well. The object SDSS J104807+005543 
is from Zhou et al. (2005). Overplotted circles are four dual AGNs known from published literature.}
\label{fig4}
}

\subsection{Kepler diagram}
The process of merging of two galaxies can be divided into three phases (Begelman et al. 
1980). At the time of merging, the bulges are homogenized within the friction timescale, 
but the narrow line regions are still bounded to their original host systems. Galaxies in 
this stage are the most plausible sites for hosting dual SMBHs and are expected to exhibit 
double nuclei in the optical and infrared. Since the \oiii$\lambda5007$ emission is the most
powerful narrow line, we will mainly use this line in the Kepler diagram.

According to Kepler's law, for a binary system with the
components $M_1$ and $M_2$, the orbital velocity ratio of the two components is given by
\begin{equation}
\frac{V_1}{V_2}= \frac{M_2}{M_1},
\end{equation}
if the total angular momentum is conserved for the circular orbital system. An emission line
with rest wavelength $\lambda_0$ will split into red and blue components due to the orbital 
rotation around the mass center, and we have
\begin{equation}
\frac{V_1}{V_2}=\frac{\Delta \lambda_1}{\Delta \lambda_2},
\end{equation}
where blue- and redshifts are measured by $\Delta \lambda_1=\left|\lambda_1-\lambda_0\right|$
and $\Delta \lambda_2=\left|\lambda_2-\lambda_0\right|$, $\lambda_1$ and $\lambda_2$ are the
wavelengths at the rest frame of the host galaxies. 

The masses of the two components are 
difficult to measure directly, but their ratio can be reasonably estimated from the 
\oiii$\lambda5007$ luminosity. For a binary system with large separation, the potential
of the two components will include mass of the bulges. However, the mass ratio should be equal 
to the SMBH ratio provided the Magorrian relation holds (Magorrian et al. 1998; Marconi \& 
Hunt 2003). We have $M_{\bullet,1}(M_{\bullet,2})\propto M_1(M_2)$, where $M_{\bullet,1}$ 
and $M_{\bullet,2}$ are the masses of the SMBH components corresponding to the two progenitor 
galaxies. If we measure the luminosities of the blue and red components of the
\oiii$\lambda5007$ line, we have
\begin{equation}
\frac{M_{\bullet,2}}{M_{\bullet,1}}=\epsilon_{1,2}\frac{L_{[\rm O~III],2}}{L_{[\rm O~III],1}},
\end{equation}
where $\epsilon_{1,2}=\epsilon_1/\epsilon_2$, $\epsilon_1$ and $\epsilon_2$ are the Eddington 
ratios of the two SMBHs (the bolometric luminosity is obtained from
$L_{\rm Bol}={\cal C}L_{\rm [O~III]}=\epsilon L_{\rm Edd}$. ${\cal C}$ is a bolometric
correction factor from the \oiii luminosity). The dynamical criterion for having dual AGNs 
is then given by
\begin{equation}
\frac{L_{[\rm O~III],1}}{L_{[\rm O~III],2}}=\epsilon_{1,2}\frac{\Delta \lambda_2}{\Delta \lambda_1}.
\end{equation}
In principle, $\epsilon_{1,2}$ is unknown observationally. Interestingly, the flux ratio 
range of the sample corresponds to the major merger systems if the double-peaked sources 
are dual AGNs and each AGN is accreting at similar Eddington ratios. Therefore 
$\epsilon_{1,2}\sim 1$ holds for the two SMBHs, which are driven by similar environments 
of tidal interaction between the two galaxies during major mergers. This is also supported
by the correlation from Fig. 4.

Equation (5) has some striking advantages for finding dual AGNs: 1) both sides are easily 
measured from observed spectra; 2) the equation is independent of the inclinations of the 
systems. We stress here that equation (5) is a dynamical relation to examine the origin 
of the double-peaked systems.

In order to understand the implication of the correlation observed (eq. 1), we need to 
compare it with eq. (5). We find that the correlation (eq. 1) agrees well with eq. (5) 
if $\epsilon_{1,2}\sim 1$ in case of ${\cal D}\sim 0$. The correlation lends strong 
support for the scenario of dual AGNs in our sample. We thus have $\sim$ 90 candidates 
from $\sim 7000$ AGNs, and the fraction of dual AGN candidates is about 
$f_{\bullet\bullet}\sim 10^{-2}$. This nicely agrees with the fraction of major
mergers in the local universe from observations (Ryan et al. 2008), but is much higher than 
the presence of binary SMBHs with a probability of $10^{-4}$ as found by Boroson \& Lauer
(2009). 

According to Kepler's law, we have a separation 
$A=0.48q^2M_{10}(1+q)^{-1}\delta_{-3}^{-2}~{\rm kpc}$, where the mass ratio $q=M_2/M_1$, 
$\delta_{\lambda_1}=|\lambda_1/\lambda_0-1|$, and $\delta_{-3}=\delta_{\lambda_1}/10^{-3}$ 
and $M_{10}=M_1/10^{10}M_\odot$ is the mass of the bulge or spheroid. Though the exact 
masses of the systems are unknown, the separation should be typically less than 1 kpc
(smaller than 1.4 kpc in NGC 6240, Komossa et al. 2003). A visual inspection of our 
candidates shows they most often have structures of tidal tails, that can be related 
to major mergers. They thus likely contain two black holes as suggested by numerical 
simulations.

The uncertainties in eq. (5) are mainly due to the uncertainties in $\epsilon_{1,2}$. 
The dispersion of the parameter $\epsilon_{1,2}$ is expressed by
$\sigma(\epsilon_{1,2})/\langle\epsilon_{1,2}\rangle=\sqrt{2}\sigma(\epsilon_1)/\langle\epsilon_1\rangle$, 
where $\langle\rangle$ indicates average values, and we use
$\sigma(\epsilon_1)/\langle\epsilon_1\rangle=\sigma(\epsilon_2)/\langle\epsilon_2\rangle$.
It turns out that $\sigma(\epsilon_1)/\langle\epsilon_1\rangle\approx 0.3$ based on local 
AGNs (Greene \& Ho 2007), and hence $\sigma(\epsilon_{1,2})/\langle\epsilon_{1,2}\rangle\approx 0.4$. 
It should be noted that the dispersion in $\epsilon_{1,2}$ is probably a lower limit on the true 
scatters in the Kepler diagram. Additionally, the actual orbit of dual AGNs could be elliptical 
rather than a perfect circular, leading to additional scatters in Fig. 4.

\subsection{Disfavoring other explanations}
Disk or bi-polar outflows are able to produce double-peaked \oiii profiles, and are generally 
employed to interpret the origin of double peaks (e.g. Eracleuous \& Halpern 2003;
Heckman 1981, 1984; Whittle 1985a,b). For bi-polar symmetric outflows, the observed fluxes of 
the blue- and redshifted components are ${\cal D}_{\rm blue}^3I_0$ and ${\cal D}_{\rm red}^3I_0$, 
where $I_0$ is the intrinsic intensity of the emission line at $\lambda_0$,
${\cal D}_{\rm blue}=1/\Gamma(1-\beta\cos\Theta)$ and ${\cal D}_{\rm red}=1/\Gamma(1+\beta\cos\Theta)$
are the Doppler factors of the approaching and receding outflows, respectively, 
$\beta=V_{\rm out}/c$ is the velocity of the outflows, $\Gamma=1/(1-\beta^2)^{1/2}$ is the Lorentz 
factor and $\Theta$ is the orientation angle of the outflow with respect to observers (e.g. Rybicki 
\& Lightman 1979). We have the luminosity ratio
$L_{[\rm O~III],1}/L_{[\rm O~III],2}=(1+\beta\cos\Theta)^3/(1-\beta\cos\Theta)^3$ whereas the shift
ratio $\Delta \lambda_1/\Delta \lambda_2=(1+\beta\cos\Theta)/(1-\beta\cos\Theta)$. It is then
expected that $L_{[\rm O~III],1}/L_{[\rm O~III],2}=\left(\Delta \lambda_1/\Delta \lambda_2\right)^3$.
This directly conflicts with the correlation discovered from the sample. A similar relation
is expected for the disk origin though there is a geometry-dependence of the emission line profiles.
Additionally the range of the ratios should be very narrow since the velocities in outflows and disk 
rotations are much smaller than the light speed. The current correlation does not favor the disk or 
bi-polar origins of the double peak profiles. 

\subsection{Additional notes}
The object SDSS J094124+394441 in our sample is 3C 223.1,
which is a well-known X-shaped radio source (Merritt \& Ekers 2002). We mark this object with a 
red circle in Fig. 4. Additionally, Cheung (2007) provides a list of 100 X-shaped radio objects:
22 of these have SDSS spectra, and one (SDSS J143017+521735) has double-peaked narrow emission 
lines (see the blue circle in Fig. 4). The two independent confirmations lend more support to  
that these two objects may be dual AGNs.

If activities of SMBHs are triggered via mergers (Di Matteo
et al. 2005), a transition phase of dual AGNs should be common before the formation of the SMBH binaries,
which are bounded by the potential of the two black holes (Begelman et al. 1980). X-shaped radio sources
have been suggested to be binary black holes (Merritt \& Ekers 2002) in light of two double-sided jets.
However, the fraction of radio-loud AGNs is $\sim 10\%$, meaning that only $\sim 1\%$ of dual AGNs
are expected to be X-shaped in the radio. Such low occurrence makes radio images relatively non-effective 
in searching for dual AGNs or SMBH binaries, but the double-peaked sources could be identified by radio 
morphology as a supplementary criterion. Multi-wavelength observations of these sources are important in
further verifying their properties.

\section{Conclusions}
We find a striking correlation between the ratios of \oiii shifts and the double peak fluxes in
a sample of 87 AGNs from the SDSS. This correlation strongly favors the explanation of dual AGNs 
in light of the Kepler diagram and does not favor a disk and outflow origin. The results also show  
that dual AGNs at 1 kpc scale are common, strongly impacting on the observational deficit of binary
SMBHs with a probability of $\sim 10^{-4}$ as shown by Boroson \& Lauer (2009).  
The present correlation in the Kepler diagram and the dynamical criterion offer an efficient way
to select targets (e.g. full version of Table 1) for future image detections in optical band and 
X-rays.

Finally, redshifts of host galaxies are a vital parameter in determination of the position of
sources in the Kepler diagram. Higher resolution spectra are extremely important to diagnose dual
AGNs among the weak double-peaked sources. The method presented in this paper is ideal for the study 
of such sources using high resolution spectra in the future.

\acknowledgements We are very grateful to the referee for useful comments and suggestions that have 
strengthened this work. T. Heckman, G. Kauffmann, C. Tremonti, R. A. Overzier, H. Netzer, T.-G. Wang 
and L. C. Ho are greatly thanked for useful discussions and comments, especially, C. Tremonti who 
helped us determine the redshifts of host galaxies. We appreciate the stimulating discussions 
among the members of the IHEP AGN group. The research is supported by NSFC-10733010 and 10821061, 
CAS-KJCX2-YW-T03, and 973 project (2009CB824800). 

{\em A note added}. We were pleased to learn from a talk presented in the KIAA meeting on July 24, 
2009 that D. Xu \& S. Komossa (see their paper: arXiv:0908.3140) were working on one individual 
double-peaked object. After we submitted our paper on 27 July, 2009, two more
papers have appeared on astro-ph (Smith et al., arXiv:0908.1998; Liu et al., arXiv:0908.2426). 
These studies are highly complementary in this hot topic.

\begin{table}
\centering 
\footnotesize
\caption{The sample of 87 type 2 AGNs}
\begin{tabular}{lcccclr}\hline\hline
Name & $z$ & $\Delta \lambda_1$ & $\Delta \lambda_2$ &$F_{\rm [O~III]1}$ &$F_{\rm [O~III]2}$  &
${\cal D}$ \\ 
(1) &(2) & (3) &(4)& (5)&(6)  &(7)\\ \hline
SDSS J000249$+$004504&    0.08662$\pm5\times 10^{-5}$&    5.18$\pm$0.07&    3.76$\pm$0.05&  662$\pm$13&  731$\pm$13&    0.13      \\
SDSS J000656$+$154847&    0.12515$\pm2\times 10^{-5}$&    3.30$\pm$0.10&    3.14$\pm$0.09&  618$\pm$19&  498$\pm$19&    0.10      \\
SDSS J013555$+$143529&    0.07208$\pm2\times 10^{-5}$&    2.35$\pm$0.08&    2.30$\pm$0.05&  586$\pm$17&  328$\pm$15&    0.14      \\
SDSS J014209$-$005049&    0.13253$\pm2\times 10^{-5}$&    1.20$\pm$0.12&    2.94$\pm$0.17&  360$\pm$28&  377$\pm$28&      ---     \\
SDSS J015605$-$000721&    0.08084$\pm2\times 10^{-5}$&    2.77$\pm$0.08&    2.67$\pm$0.10&  545$\pm$18&  427$\pm$18&      ---     \\
SDSS J073509$+$403624&    0.10297$\pm2\times 10^{-5}$&    1.99$\pm$0.11&    2.35$\pm$0.04&  686$\pm$25&  640$\pm$24&      ---     \\
SDSS J074729$+$344018&    0.12982$\pm2\times 10^{-5}$&    3.27$\pm$0.10&    2.11$\pm$0.13&   57$\pm$ 4&   69$\pm$ 4&      ---     \\
SDSS J074953$+$451454&    0.03132$\pm1\times 10^{-5}$&    1.97$\pm$0.09&    2.35$\pm$0.07&  746$\pm$31&  796$\pm$31&      ---     \\
SDSS J075223$+$273643&    0.06908$\pm2\times 10^{-5}$&    2.24$\pm$0.05&    2.42$\pm$0.03&  747$\pm$16&  980$\pm$16&    0.03      \\
SDSS J080218$+$304622&    0.07654$\pm3\times 10^{-5}$&    2.99$\pm$0.04&    3.02$\pm$0.04& 2002$\pm$31& 2575$\pm$32&    0.02      \\
SDSS J080418$+$305157&    0.14553$\pm2\times 10^{-5}$&    3.25$\pm$0.06&    4.83$\pm$0.03&  210$\pm$ 4&  198$\pm$ 4&    0.61      \\
SDSS J080740$+$390015&    0.02346$\pm1\times 10^{-5}$&    5.44$\pm$0.06&    4.72$\pm$0.07& 1762$\pm$26& 1529$\pm$25&      ---     \\
SDSS J080841$+$481351&    0.12365$\pm4\times 10^{-5}$&    4.82$\pm$0.16&    1.85$\pm$0.12&  414$\pm$23&  654$\pm$23&    0.18      \\
SDSS J081430$+$265729&    0.07874$\pm4\times 10^{-5}$&    3.04$\pm$0.06&    2.22$\pm$0.06&  430$\pm$15&  593$\pm$16&    0.09      \\
SDSS J082107$+$502115&    0.09534$\pm3\times 10^{-5}$&    2.85$\pm$0.16&    2.85$\pm$0.09&  281$\pm$15&  312$\pm$14&      ---     \\
SDSS J085416$+$502631&    0.09553$\pm4\times 10^{-5}$&    2.32$\pm$0.04&    3.16$\pm$0.03&  502$\pm$ 9&  520$\pm$ 9&    0.25      \\
SDSS J085512$+$642345&    0.03621$\pm1\times 10^{-5}$&    2.52$\pm$0.07&    2.26$\pm$0.06&  372$\pm$11&  338$\pm$11&      ---     \\
SDSS J085841$+$104122&    0.14810$\pm5\times 10^{-5}$&    3.31$\pm$0.05&    3.25$\pm$0.04&  863$\pm$11&  653$\pm$11&    0.69      \\
SDSS J094124$+$394441&    0.10750$\pm3\times 10^{-5}$&    2.68$\pm$0.08&    2.10$\pm$0.07&  774$\pm$32& 1218$\pm$33&    0.29      \\
SDSS J094427$+$144717&    0.07744$\pm2\times 10^{-5}$&    4.59$\pm$0.12&    1.86$\pm$0.09&   66$\pm$ 5&  194$\pm$ 6&      ---     \\
SDSS J095528$+$383827&    0.06124$\pm2\times 10^{-5}$&    3.38$\pm$0.14&    4.45$\pm$0.20&  309$\pm$14&  252$\pm$12&    0.70      \\
SDSS J095833$-$005118&    0.08587$\pm3\times 10^{-5}$&    3.23$\pm$0.08&    3.56$\pm$0.05&  674$\pm$17&  785$\pm$17&    0.23      \\
SDSS J101143$+$325943&    0.12974$\pm4\times 10^{-5}$&    2.71$\pm$0.02&    3.26$\pm$0.02&  974$\pm$10& 1220$\pm$11&    0.03      \\
SDSS J101346$-$005451&    0.04243$\pm2\times 10^{-5}$&    4.57$\pm$0.12&    3.02$\pm$0.08&  359$\pm$10&  574$\pm$11&      ---     \\
SDSS J101348$+$002014&    0.11711$\pm3\times 10^{-5}$&    2.39$\pm$0.33&    3.08$\pm$0.14&  156$\pm$17&  178$\pm$16&      ---     \\
SDSS J102325$+$324348&    0.12701$\pm5\times 10^{-5}$&    3.04$\pm$0.20&    3.41$\pm$0.15&  189$\pm$14&  244$\pm$14&    0.39      \\
SDSS J104813$+$442710&    0.13972$\pm3\times 10^{-5}$&    3.13$\pm$0.27&    2.47$\pm$0.14&   51$\pm$ 6&   80$\pm$ 6&      ---     \\
SDSS J105653$+$331945&    0.05104$\pm2\times 10^{-5}$&    3.75$\pm$0.04&    5.38$\pm$0.07&  944$\pm$11&  553$\pm$ 9&      ---     \\
SDSS J110215$+$290725&    0.10603$\pm3\times 10^{-5}$&    3.08$\pm$0.13&    4.22$\pm$0.11&  295$\pm$12&  229$\pm$11&    0.04      \\
SDSS J110821$+$591851&    0.08572$\pm2\times 10^{-5}$&    2.48$\pm$0.05&    2.49$\pm$0.05&  919$\pm$21&  968$\pm$21&    0.04      \\
SDSS J110832$+$195128&    0.10317$\pm3\times 10^{-5}$&    3.54$\pm$0.22&    3.19$\pm$0.12&  325$\pm$26&  692$\pm$27&    0.27      \\
SDSS J110957$+$020138&    0.06322$\pm2\times 10^{-5}$&    3.06$\pm$0.07&    2.85$\pm$0.03&  723$\pm$21& 1863$\pm$26&    0.17      \\
SDSS J111042$+$030033&    0.10752$\pm4\times 10^{-5}$&    2.18$\pm$0.08&    3.46$\pm$0.09&  302$\pm$11&  356$\pm$11&    0.47      \\
SDSS J111054$+$012936&    0.12946$\pm4\times 10^{-5}$&    1.52$\pm$0.08&    2.37$\pm$0.06&  526$\pm$22&  633$\pm$22&      ---     \\
SDSS J111201$+$275053&    0.04743$\pm2\times 10^{-5}$&    3.85$\pm$0.18&    2.18$\pm$0.12&  229$\pm$18&  435$\pm$19&    0.45      \\
SDSS J111333$+$165711&    0.05343$\pm2\times 10^{-5}$&    1.68$\pm$0.12&    4.46$\pm$0.04&  293$\pm$ 8&  380$\pm$ 9&    0.41      \\
SDSS J111803$+$062657&    0.12781$\pm2\times 10^{-5}$&    2.73$\pm$0.15&    3.72$\pm$0.14&  311$\pm$16&  298$\pm$16&    0.50      \\
SDSS J112659$+$294442&    0.10180$\pm3\times 10^{-5}$&    2.11$\pm$0.03&    3.18$\pm$0.04&  843$\pm$12&  680$\pm$12&    0.24      \\
SDSS J113126$-$020459&    0.14641$\pm5\times 10^{-5}$&    4.04$\pm$0.05&    2.66$\pm$0.06&  452$\pm$ 9&  320$\pm$ 8&      ---     \\
SDSS J113630$+$135848&    0.08173$\pm3\times 10^{-5}$&    3.05$\pm$0.18&    1.07$\pm$0.15&   74$\pm$11&  190$\pm$12&      ---     \\
SDSS J113759$+$143733&    0.08103$\pm2\times 10^{-5}$&    2.74$\pm$0.14&    2.54$\pm$0.11&  270$\pm$18&  401$\pm$18&      ---     \\
SDSS J114047$+$445208&    0.06917$\pm2\times 10^{-5}$&    5.00$\pm$0.10&    1.91$\pm$0.20&  118$\pm$ 6&  149$\pm$ 7&      ---     \\
SDSS J114249$+$102700&    0.11870$\pm4\times 10^{-5}$&    3.15$\pm$0.24&    3.58$\pm$0.09&  110$\pm$ 6&  136$\pm$ 6&      ---     \\
SDSS J114610$-$022619&    0.12251$\pm4\times 10^{-5}$&    2.23$\pm$0.04&    1.95$\pm$0.06&  348$\pm$10&  421$\pm$10&      ---     \\
SDSS J114840$-$021637&    0.08481$\pm3\times 10^{-5}$&    3.21$\pm$0.05&    2.97$\pm$0.07&  406$\pm$11&  491$\pm$12&    0.28      \\
SDSS J115028$+$044141&    0.04037$\pm1\times 10^{-5}$&    2.74$\pm$0.08&    2.55$\pm$0.04& 1128$\pm$36& 2300$\pm$39&    0.05      \\
SDSS J115249$+$190300&    0.09666$\pm4\times 10^{-5}$&    2.25$\pm$0.12&    3.12$\pm$0.05&  700$\pm$23&  561$\pm$22&      ---     \\
SDSS J115548$+$110159&    0.10506$\pm5\times 10^{-5}$&    3.21$\pm$0.23&    1.70$\pm$0.29&   83$\pm$11&   86$\pm$11&      ---     \\
SDSS J120320$+$131931&    0.05844$\pm2\times 10^{-5}$&    2.57$\pm$0.14&    2.37$\pm$0.13&  172$\pm$13&  193$\pm$13&      ---     \\
SDSS J120357$+$280856&    0.06934$\pm2\times 10^{-5}$&    2.21$\pm$0.10&    2.40$\pm$0.11&  208$\pm$10&  140$\pm$ 9&    0.89      \\
SDSS J120802$+$230430&    0.07363$\pm3\times 10^{-5}$&    2.41$\pm$0.09&    2.76$\pm$0.12&  263$\pm$11&  188$\pm$11&    0.03      \\
SDSS J121311$+$650818&    0.13541$\pm4\times 10^{-5}$&    1.84$\pm$0.14&    2.79$\pm$0.05&   54$\pm$ 4&  177$\pm$ 5&    0.06      \\
SDSS J121527$+$000109&    0.10033$\pm3\times 10^{-5}$&    2.12$\pm$0.42&    1.78$\pm$0.42&  196$\pm$49&  244$\pm$49&      ---     \\
SDSS J123524$+$060810&    0.09413$\pm2\times 10^{-5}$&    3.13$\pm$0.12&    2.57$\pm$0.07&  270$\pm$13&  425$\pm$13&      ---     \\
SDSS J123538$+$050221&    0.06532$\pm2\times 10^{-5}$&    2.01$\pm$0.29&    2.36$\pm$0.13&   65$\pm$ 9&   88$\pm$ 9&      ---     \\
SDSS J125049$+$074618&    0.05038$\pm2\times 10^{-5}$&    1.52$\pm$0.08&    2.19$\pm$0.09&  195$\pm$10&  144$\pm$ 9&      ---     \\
SDSS J131235$+$500415&    0.11586$\pm3\times 10^{-5}$&    2.79$\pm$0.19&    3.72$\pm$0.22&  198$\pm$17&   87$\pm$12&      ---     \\
SDSS J131515$+$213403&    0.07028$\pm3\times 10^{-5}$&    4.58$\pm$0.08&    2.16$\pm$0.05&  863$\pm$22& 1364$\pm$23&    0.27      \\
SDSS J132547$+$545019&    0.14183$\pm4\times 10^{-5}$&    1.76$\pm$0.12&    2.76$\pm$0.16&  396$\pm$27&  412$\pm$27&      ---     \\
SDSS J133723$+$035350&    0.02280$\pm1\times 10^{-5}$&    2.42$\pm$0.08&    2.75$\pm$0.16&  514$\pm$23&  272$\pm$22&    0.03      \\
SDSS J133804$+$264209&    0.10667$\pm2\times 10^{-5}$&    1.92$\pm$0.56&    1.83$\pm$0.40&   67$\pm$18&   55$\pm$18&      ---     \\
SDSS J135207$+$052555&    0.07883$\pm2\times 10^{-5}$&    2.14$\pm$0.15&    3.15$\pm$0.05&  714$\pm$33&  952$\pm$32&    0.01      \\
SDSS J142039$+$495906&    0.06333$\pm2\times 10^{-5}$&    2.74$\pm$0.19&    1.80$\pm$0.10&  165$\pm$16&  318$\pm$17&      ---     \\
SDSS J143132$+$435807&    0.09604$\pm4\times 10^{-5}$&    2.89$\pm$0.33&    3.23$\pm$0.35&   60$\pm$ 9&   66$\pm$ 9&      ---     \\
SDSS J143434$+$140548&    0.06974$\pm3\times 10^{-5}$&    1.99$\pm$0.36&    3.21$\pm$0.44&  140$\pm$22&  119$\pm$22&      ---     \\ \hline
\end{tabular}
\end{table}

\newpage
\begin{table*}
\centering
{{\sc Table 1}--- {\em Continued}}\\
\vglue 0.2cm
\footnotesize
\begin{tabular}{lcccclr}\hline\hline
Name & $z$ & $\Delta \lambda_1$ & $\Delta \lambda_2$ &$F_{\rm [O~III]1}$ &$F_{\rm [O~III]2}$  & ${\cal D}$ \\ 
(1) &(2) & (3) &(4)& (5)&(6)  &(7)\\ \hline
SDSS J145217$+$511050&    0.07582$\pm3\times 10^{-5}$&    2.08$\pm$0.17&    3.79$\pm$0.12&  500$\pm$25&  379$\pm$24&      ---     \\
SDSS J145801$+$274251&    0.11845$\pm3\times 10^{-5}$&    2.79$\pm$0.23&    2.61$\pm$0.26&  207$\pm$20&  201$\pm$21&      ---     \\
SDSS J150053$+$382349&    0.14523$\pm5\times 10^{-5}$&    3.10$\pm$0.08&    2.59$\pm$0.09&  475$\pm$15&  452$\pm$15&    0.34      \\
SDSS J150452$+$321414&    0.11253$\pm4\times 10^{-5}$&    1.86$\pm$0.21&    2.38$\pm$0.12&  194$\pm$21&  277$\pm$25&      ---     \\
SDSS J151659$+$051751&    0.05122$\pm1\times 10^{-5}$&    2.10$\pm$0.04&    3.31$\pm$0.03&  685$\pm$11&  729$\pm$12&    0.36      \\
SDSS J151709$+$335324&    0.13541$\pm6\times 10^{-5}$&    6.75$\pm$0.01&    5.57$\pm$0.02& 3659$\pm$16& 2232$\pm$11&      ---     \\ 
SDSS J152606$+$414014&    0.00831$\pm0.7\times 10^{-5}$&  2.92$\pm$0.04&    1.83$\pm$0.05& 3146$\pm$96& 6776$\pm$105&    0.15      \\
SDSS J153217$+$453225&    0.07138$\pm3\times 10^{-5}$&    2.71$\pm$0.12&    1.58$\pm$0.07&  170$\pm$11&  261$\pm$11&    0.14      \\
SDSS J153315$+$575001&    0.06734$\pm2\times 10^{-5}$&    3.08$\pm$0.06&    2.37$\pm$0.09&  337$\pm$11&  311$\pm$11&      ---     \\
SDSS J155205$+$043317&    0.08016$\pm3\times 10^{-5}$&    2.78$\pm$0.09&    2.14$\pm$0.05&  112$\pm$ 6&  298$\pm$ 6&      ---     \\
SDSS J155619$+$094855&    0.06785$\pm1\times 10^{-5}$&    3.12$\pm$0.04&    3.65$\pm$0.02&  783$\pm$10& 1112$\pm$10&    0.11      \\
SDSS J160024$+$264035&    0.09003$\pm4\times 10^{-5}$&    3.94$\pm$0.36&    3.48$\pm$0.26&   88$\pm$12&  153$\pm$13&    0.11      \\
SDSS J160436$+$500958&    0.14649$\pm3\times 10^{-5}$&    2.94$\pm$0.04&    3.22$\pm$0.05&  353$\pm$ 7&  354$\pm$ 7&    0.08      \\
SDSS J160524$+$152233&    0.04210$\pm1\times 10^{-5}$&    3.23$\pm$0.18&    2.81$\pm$0.12&  200$\pm$17&  386$\pm$17&    0.37      \\
SDSS J160636$+$342754&    0.05473$\pm2\times 10^{-5}$&    2.70$\pm$0.20&    1.94$\pm$0.17&  230$\pm$23&  288$\pm$23&      ---     \\
SDSS J161006$+$210735&    0.13722$\pm2\times 10^{-5}$&    3.49$\pm$0.05&    2.01$\pm$0.03&  170$\pm$ 5&  486$\pm$ 6&    0.21      \\
SDSS J162939$+$240856&    0.05921$\pm2\times 10^{-5}$&    2.60$\pm$0.13&    1.69$\pm$0.13&  133$\pm$10&  115$\pm$ 9&      ---     \\
SDSS J163316$+$262716&    0.07125$\pm4\times 10^{-5}$&    2.10$\pm$0.21&    3.11$\pm$0.29&  160$\pm$15&  105$\pm$15&      ---     \\
SDSS J165206$+$310707&    0.07495$\pm2\times 10^{-5}$&    2.04$\pm$0.08&    3.15$\pm$0.07&  115$\pm$ 5&  157$\pm$ 5&    0.09      \\
SDSS J225420$-$005134&    0.07950$\pm3\times 10^{-5}$&    3.17$\pm$0.11&    3.78$\pm$0.12&  183$\pm$ 8&  112$\pm$ 7&    0.15      \\
SDSS J230442$-$093345&    0.03203$\pm1\times 10^{-5}$&    2.69$\pm$0.10&    2.67$\pm$0.17&  449$\pm$24&  393$\pm$25&      ---     \\
SDSS J231051$-$090011&    0.09439$\pm2\times 10^{-5}$&    1.20$\pm$0.03&    4.28$\pm$0.03&  640$\pm$ 8&  428$\pm$ 7&    0.20      \\ \hline
\end{tabular}
\parbox{5.25in}{\baselineskip 10.5pt
\noindent
{\sc Note}:
Col (1): SDSS names of objects; Col (2)  redshifts and their uncertainties;
Col. (3) and (4) are the Doppler red and blue shifts of \oiii\, line in units of \AA,
respectively.; Col (5) and (6) are the corresponding fluxes in units of
$10^{-17}$ergs~s$^{-1}$~cm$^{-2}$, respectively. (7) parameter ${\cal D}$ is the
separation of the two components in the BPT diagram.\\
This is only part of the full table. See the complete sample online.}
\end{table*}
\end{document}